\documentclass[british]{article}
\usepackage[T1]{fontenc}
\usepackage[utf8]{inputenc}
\usepackage{lmodern}
\usepackage[a4paper]{geometry}
\usepackage{babel}
\usepackage{amsmath}
\usepackage{algorithm}
\usepackage[noend]{algpseudocode}
\usepackage{amssymb}
\usepackage{cite}
\usepackage{hyperref}
\usepackage{graphicx}
\usepackage{subcaption}
\usepackage{tikz}
\usepackage{ragged2e}
\usepackage{authblk}
\usepackage{amsthm}

\newtheorem{definition}{Definition}
\newtheorem{assumption}{Assumption}
\usepackage[T1]{fontenc}
\usepackage[utf8]{inputenc}
\usepackage{authblk}

\renewrobustcmd{\bfseries}{\fontseries{b}\selectfont}
\renewrobustcmd{\boldmath}{}
\newrobustcmd{\B}{\bfseries}

\title{Strong Transitivity Relations and Graph Neural Networks}

\date{}

\author[1]{Yassin Mohamadi\thanks{yassinmohamadi@aut.ac.ir}}
\author[1]{Mostafa Haghir Chehreghani\thanks{mostafa.chehreghani@aut.ac.ir (corresponding author)}}

\affil[1]{Department of Computer Engineering, Amirkabir University of Technology (Tehran Polytechnic), Tehran, Iran}

\begin{document}
	
\maketitle

\begin{center}
	Abstract\\
	
	\justifying
Local neighborhoods play a crucial role in embedding generation in graph-based learning. It is commonly believed that nodes ought to have embeddings that resemble those of their neighbors. In this research, we try to carefully expand the concept of similarity from nearby neighborhoods to the entire graph. We provide an extension of similarity that is based on transitivity relations, which enables Graph Neural Networks (GNNs) to capture both global similarities and local similarities over the whole graph. We introduce Transitivity Graph Neural Network (TransGNN), which more than local node similarities,
takes into account global similarities by
distinguishing strong transitivity relations from weak ones and exploiting them.
We evaluate our model over several real-world datasets and showed that it considerably improves the performance of several well-known GNN models, for tasks such as node classification.
\end{center}


\section{Intoduction}
\label{Intoduction}

Graph neural networks (GNNs), pioneered by Gori et al. ~\cite{gori2005new} and Scarselli et al. \cite{scarselli2008graph}, have experienced a surge in popularity, inspired by subsequent works by Duvenaud et al.~\cite{duvenaud2015convolutional}, Atwood and Towsley~\cite{atwood2016diffusion}, Bronstein et al.~\cite{bronstein2017geometric}, and Monti et al.~\cite{monti2017geometric}. This popularity can be attributed to GNNs' adaptability and efficiency in learning from data structured as graphs, proving essential in domains where data can be naturally organized into nodes, and predictions rely on the complex relationships (edges) inter-linking these nodes. Their versatility finds applications in diverse fields such as molecular chemistry~\cite{liao2019lanczosnet}, social networks~\cite{chen2018fastgcn}, and recommendation systems~\cite{gao2023survey}.

Graph Convolutional Networks (GCNs)~\cite{kipf2016semi}, introduced by Kipf and Welling in 2017, present an efficient adaptation of Convolutional Neural Networks (CNNs)~\cite{chua1993cnn} for graph data. This model involves stacking layers of first-order spectral filters, succeeded by a non-linear activation function, facilitating the acquisition of graph representations~\cite{kipf2016semi}.
Within the GNN framework, the core concept revolves around iteratively updating node states through interactions with their neighbors. Different GNN iterations, proposed by Wu et al.~\cite{wu2019simplifying}, Xu et al.~\cite{xu2018powerful}, Li et al.~\cite{li2015gated}, and others, present distinct strategies for aggregating neighboring representations alongside their own.
One of the key developments in this field is the Graph Attention Network (GAT), proposed by Velickovic et al.~\cite{velickovic2017graph}
Instead of using conventional methods like averaging or max-pooling to aggregate information from neighbors of each node in a graph, GAT allows each node to selectively focus on the most relevant neighbors for its representation update, thus transforming the way information is processed in graphs.
This method is further refined by efforts such as GATv2~\cite{brody2021attentive}.

The Simplifying Graph Convolutional Network (SGC)~\cite{wu2019simplifying}
leverages a simplified approach, emphasizing the importance of model simplicity and interpretability in GNNs~\cite{DBLP:journals/natmi/Chehreghani22}. It uses a novel Laplacian graph convolution operation that streamlines GCNs. This approach simplifies spectral convolutions by approximating them with a more computationally efficient quadratic formulation. SGC achieves comparable performance to traditional GCNs while reducing computational complexity and parameter count~\cite{wu2019simplifying}.
The Graph Isomorphism Network (GIN)~\cite{xu2018powerful} 
addresses graph isomorphism through a deep neural network architecture that iteratively updates node representations while capturing global graph features. 

All of the previously discussed GNNs focus on learning embeddings for nodes, edges, or graphs by leveraging the {\em presence} of connections between nodes, which means they consider one level (and an specific type of) similarity in the graph. In our exploration, the contributions of \cite{morris2019weisfeiler,damke2020supplementary} are instrumental in the introduction of a collection of GNNs, rooted in the extension of the Weisfeiler-Leman test~\cite{weisfeiler1968reduction}. Their primary endeavor centers on broadening the network of connections between nodes, extending the boundaries beyond immediate neighbors to encompass non-immediate neighbors. However, their approach predominantly emphasizes a particular type of similarity, computed based on
a local neighborhood of each node, referred as local neighborhood similarity.

In this study, our primary aim is to explore a different category of connections and similarities that extend beyond the realm of local similarities. Drawing from the concepts of transitivity and counter-similarity, we introduce a new dimension to similarity in GNNs that reaches beyond neighborhood-based similarities. As a result, we introduce two distinct levels of transitivity within a graph: strong transitivity and weak transitivity. Strong transitivity is characterized by the presence of numerous connecting paths between two nodes, while weak transitivity is observed when there are few connecting paths between the  nodes. This categorization allows us to capture and distinguish various types of transitivity relations in the graph.
In this paper, we demonstrate that strong transitivity holds promise for expanding the notion of node similarity within graph neural networks.
It captures global similarities that permeate the entire graph within GNNs.
Building upon the basic assumption that connected nodes tend to share labels and embeddings, we extend standard loss functions to explore how strong transitivity relations influence label sharing and embedding similarity between nodes.
To simultaneously incorporate both real connections (edges) and strong transitivities,
we employ a bipartite GNN that focuses on real edges and strong transitivity relations.
We create a transitivity graph based on transitivity relations, where the edges in the transitivity graph reflect the transitivity relations in the original graph.
To extract strong transitivity relations from the transitivity graph,
we apply a clustering algorithm with the objective of minimizing inter-cluster edges and maximizing intra-cluster edges.
In this way, our bipartite GNN model 
is augmented with both the original graph and the transitive graph\footnote{The implementation of our proposed method is publicly available at\url{https://github.com/yassinmihemedi/Strong-transitivity-relations-and-Graph-neural-network}}.

Our research offers some advancements in the field of graph neural networks:
\begin{itemize}
\item 
Extension of the concept of node similarity: the concept of node similarity is carefully expanded to include and span the full graph. With this expansion, we may obtain a more thorough knowledge of the graph by capturing and utilizing structural patterns and relationships on a global scale.
We present a novel GNN that takes into account both non-local and local neighborhood similarities. Based on strong transitivity and local similarity, this method improves the network's capacity to extract and utilize both global and local structural information in the graph, hence enhancing its performance in a variety of graph-based tasks.
\item 
Transitivity levels: In our work, we distinguish between two types of transitivity: strong and weak.
We argue that only strong transitivity relations contribute to nodes' similarity and as a result, to embeddings' similarity.
\end{itemize}
We perform extensive experiments over several real-world datasets and show that our model
can be successfully aligned with different GNN models,
namely GCN \cite{kipf2016semi}, GAT \cite{velickovic2017graph}, GATv2  \cite{brody2021attentive} and SGN \cite{wu2019simplifying}, to improve their performance.
Moreover, our method makes basic GNN models more resilient to noise.

The rest of this paper is organized as follows.
In Section~\ref{Related work},
we provide an overview on related work.
In Section~\ref{Preliminaries},
we present preliminaries and definitions used in the paper.
In Section~\ref{Method}, we present our transitivity-aware graph neural model.
We empirically evaluate the performance of our proposed model in
Section~\ref{Experiments}.
Finally, the paper is concluded in Section~\ref{Conclusion}.

\section{Related work}
\label{Related work}

Graph neural networks (GNNs) have gained significant attention in recent years due to their capability to model and analyze graph-structured data. The field of GNNs has seen substantial growth, with various architectural advancements and applications emerging.
In one of early attempts, Kipf and Max Welling \cite{kipf2016semi}
introduce the concept of Graph Convolutional Networks (GCNs).
GCNs
try to capture complex relationships within graphs through spectral graph convolutional operations.
However, they suffer from the scalability issue when dealing with large datasets due to the computational complexity of spectral-based convolutions \cite{wu2019simplifying}.
The graph convolution process is streamlined by Simplified Graph Convolutional Networks (SGCs), which enhances their efficiency and preserves their expressive power \cite{wu2019simplifying}.
Graph Attention Networks (GATs),
introduced by Velickovic et al. \cite{velickovic2017graph},
provide a novel approach to capture local patterns through adaptive attention mechanisms. GATs use attention coefficients to weigh the importance of neighboring nodes during information aggregation.
Graph Attention Networks version2 (GATv2) further advance attention mechanisms for improved neighborhood aggregation \cite{brody2021attentive}.
It introduces a more flexible mechanism, improving its capability to capture complex relationships within graphs while allowing for better scalability. This flexibility in attention mechanisms enables GATv2 to adaptively emphasize the contributions of different neighbors during aggregation.

The expressive power of GNNs received a significant boost with the emergence of Graph Isomorphism Networks (GINs) by Xu et al. \cite{xu2018powerful}.
In this work, the authors argue that a GNN achieves its maximal power
when it is equivalent to the Weisfeiler-Lehman (WL) test \cite{weisfeiler1968reduction}. They prove that a GNN reaches this level if and only if its function that maps a multiset of previous-layer embeddings to a single embedding is injective.
For more comprehensive reviews on graph neural networks,
interested readers may refer to~\cite{DBLP:journals/tnn/WuPCLZY21,DBLP:journals/natmi/Chehreghani22}.

Another related line of research pertains to graph matching, a fundamental problem in graph theory. Graph matching involves establishing correspondence between nodes in two graphs \cite{li2019graph}. Siamese networks, initially popularized for image similarity tasks \cite{bromley1993signature}, are now adapted for graph matching. 
Siamese networks learn to project graph representations into a shared space, where distances correspond to the similarity or dissimilarity between graphs \cite{ktena2018metric,ma2019deep}.
Moreover, research in graph similarity computation has advanced the understanding of how to quantify the similarity between two graphs. Siamese networks, in this context, is employed to learn similarity metrics for graphs \cite{ktena2018metric}. These networks learn to produce embeddings that represent graph structures, allowing for direct computation of similarity scores \cite{liu2019community}.

\section{Preliminaries}
\label{Preliminaries}

A graph can be formally defined as a tuple $G = (V,E,X)$, where $V$, $E$ and $X$
are respectively the set of nodes, the set of edges and nodes' feature vectors.
We can represent the nodes, edges and nodes' features by matrices, where $A\in  {(0,1)}^{|V|\times|V|}$ is the adjacency matrix, and $X\in R^{|V|\times d}$ is the feature matrix.
A (single-layer) graph neural network is a function $F$ that maps a graph with feature matrix $X$ and adjacency matrix $A$ to a new feature matrix $Z$.
In other words, $F: (A,X) \rightarrow Z$, where $Z\in R^{|V|\times d'}$.
Let $H^l\in R^{|V|\times d_l}$ be the matrix of nodes' representations (embeddings)
at layer $l$, ${h_i^l}$ be the representation vector
of node $i$ at layer $l$, and $d_l$ be its corresponding dimension.
We can define the update rule of GNNs at each layer as follows:
\begin{equation}
	\label{eq:message}
	h_i^{l+1} = \sigma\left(\verb|sum|_{j \in N(i)}\left(W^l h_j^l + b^l\right)\right),
\end{equation}
where $N(i)$ denotes the set of neighbors of node $i$, $\sigma$ is an activation function (e.g., ReLU, sigmoid), and ${W^l}$ and ${b^l}$ are respectively the trainable weight matrix and the bias vector at layer $l$. The $\verb|sum|$ operator can be replaced with any other aggregation functions, such as $\verb|max|$ pooling, $\verb|mean|$ pooling, or an attention mechanism~\cite{hamilton2017inductive}. Equation \ref{eq:message} is also known as "message passing".

\section{Method}
\label{Method}

In this section, we go through the details of our proposed method to improve graph neural networks.
First, we introduce the intuition behind our method. Then, we discuss its different components, including the construction of the transitivity graph and loss functions.

\subsection{Motivation}
\label{motivation}


In the classification of nodes of a graph, a common and reasonable assumption is that
two nodes $i$ and $j$ share the same label if they are structurally  similar. 
We can concentrate on a simple notion of node similarity in the graph space and 
assume that two nodes $i$ and $j$ are similar if and only if they are connected by an edge.
As a result, we can have this following assumption:
\begin{assumption}\label{as:astrans}
	If nodes $i$ and $j$ are connected, they will have the same label.
\end{assumption}
When dealing with a connected undirected graph,
Assumption \ref{as:astrans} implies that all nodes of the graph would possess the same label.
Since this is not the case for real-world graphs,
Assumption \ref{as:astrans} is not totally valid.
As a result, two important questions arise, are:
\begin{itemize}
\item 
Question 1.
When does Assumption \ref{as:astrans} hold?
\item 
Question 2. How can we leverage Assumption \ref{as:astrans} to enhance GNN performance in cases where it holds?
\end{itemize}

To answer Question 1, first we assume that {\em similarity} satisfies the transitivity property, which propagates {\em similarity} and as a result co-labeling in the graph.
More precisely, if node $i$ is similar to node $j$ and node $j$ is similar to node $k$,
then node $i$ becomes similar to node $k$.
Furthermore, nodes $i$ and $j$ and $k$ find the same label.
This transitivity property, in its naive form, leads to all nodes finding the same label. 
To address this issue, we consider two levels of transitivity:
strong transitivity and weak transitivity;
and assume that only strong transitivity leads to label propagation.
On the other hand, edges that reflect weak transitivity do not
significantly contribute in label propagation.

\begin{definition}
We define there is a strong transitivity between two nodes in a graph, if these two conditions are met:
i) there is a significant number of paths connecting them, and 
ii) they share similar structural characteristics.
For the sake of simplicity, we refer to the first condition
as the "Significant Number of Paths Connecting" condition, or SNPC in short;
and to the second condition as the "Similar Structural Characteristics" condition
or SSC in short.
Conversely, two nodes demonstrate weak transitivity when there are few or no paths connecting them, or they are structurally dissimilar.
\end{definition}

To answer the second question, we begin by examining Equation \ref{eq:message}.
According to Equation \ref{eq:message}, each node's embedding is influenced by its local neighbors.
To improve the performance of GNNs, it is crucial to take into account both local and global similarities within the network. Local similarity pertains to how nodes are similar to their immediate neighbors, while global similarity involves how nodes exhibit similarity across the entire graph due to strong transitivity principles.
Therefore to do so, first we extract strong transitivity relations.
Next, we incorporate both local edge information and global strong transitivity into GNN models. The foundation of GNNs is based on the idea that connected nodes in the graph should find similar embeddings.
We seek to expand this idea by examining nodes related by strong transitivity also have shared labels or embeddings.
Our idea to find strong transitivity relations in the graph
is based on building a transitivity graph and clustering it.

 \subsection{Transitivity graph construction}
\label{transitive graph}

In this section, we elucidate the process of creating the transitivity graph and address any potential challenges that may occur during this procedure.
As previously discussed, in our efforts to improve the performance of GNNs, we aim to extract strong transitivity relations from the original graph.
This modified graph is referred to as the "transitivity graph".
Before delving into our method for extracting transitivity relations from the original graph, it is essential to elucidate the rationale behind the creation of the transitivity graph.
Our first objective is to harness the combined potential of
both real connections between nodes and transitivity relations that exists among them.
Secondly, if we were to forgo the creation of the transitivity graph and instead directly apply the transitivity concept to the original graph, this approach would not only entail a straightforward structural expansion but could also exert a detrimental influence on the genuine connections or edges within the original graph. This distinction becomes evident when considering Equation \ref{eq:message}. It suggests that when we combine transitivity connections with real edges in the original graph, they appear to have the same impact on the embedding. However, it's clear that the influence of real connections should be distinct from that of transitivity connections.

To extract strong transitivity relations,
we rely on the construction a transitivity graph.
This graph must satisfy both the SSC condition and the SNPC condition:
i) the transitivity graph should aptly represent node similarities, and
ii) it should effectively convey strong transitivity relations from the original graph.
As previously highlighted, in order to prevent interference with the genuine connections and strong transitivity relations, it is imperative that the edges/connections existing in  the original graph are absent in the transitivity graph.
Edges within the transitivity graph should reflect strong transitivity relations in the original graph.
As a result, nodes are identical in both the original and transitivity graphs,
but nodes that are connected by an edge in the original graph should not be connected in the transitivity graph.
In essence, the edges present in the original graph should be deliberately excluded from the transitivity graph.
Altogether, we build the transitivity graph as follows:
\begin{itemize}
\item 	
Its nodes are the same as the nodes of the original graph.
\item 
Two nodes are connected in the transitivity graph
if all the following conditions are satisfied:
i) they exhibit a level of structural similarity (e.g., SimRank score) in the original graph
exceeding a defined minimum threshold, and
ii) they are not connected to each other in the original graph. 
\end{itemize}

It is easy to see that in this way, condition SSC is satisfied.
To additionally meet the SNPC condition,
we apply a clustering algorithm to the transitivity graph so that while its intra-cluster edges are preserved,
its inter-cluster edges are eliminated.
In the following, we explain why this heuristic method
keeps strong transitivity relations
and discards weak transitivity relations.
Our argument is based this reasonable assumption that
if nodes $v$ and $w$ belong to the same cluster
$v$ is connected to $w$'s neighbors;
and if $v$ and $u$ do not belong to the same cluster,
$v$ is not connected to $u$'s neighbors.

Assume that in the transitivity graph, node $v$ is connected to nodes $w$ and $u$,
such that $v$ and $w$ belong to the same cluster but $v$ and $u$ do not.
As a result and by clustering the graph,
while the link between $v$ and $w$, i.e., $\{v,w\}$, is preserved,
the link between $v$ and $u$, i.e., $\{v,u\}$, is discarded.
We argue that $\{v,w\}$ represents a strong transitivity relation,
whereas $\{v,u\}$ indicates a weak transitivity relation.
Therefore by clustering the transitivity graph, 
weak transitivity relations of the original graph are eliminated from the transitivity graph.
Since the number of paths contribute positively to 
most of structural similarity measures
(SimRank \cite{jeh2002simrank}, personalized PageRank  \cite{DBLP:journals/im/FogarasRCS05,DBLP:journals/widm/Chehreghani21}etc),
two nodes that find a structural similarity above a threshold are likely to have a certain number of paths connecting them.
As a result, each edge in the transitivity graph
implies the existence of a certain number of paths between its endpoints in the original graph.
Let $w_1, \ldots, w_c$ be the neighbors of $w$,
and $u_1, \ldots, u_{c'}$ be the neighbors of $u$.
Since $v$ and $w$ belong to the same cluster,
$v$ has edges to $w_1, \ldots, w_c$.
On the other hand, since $v$ and $u$ do not belong to the same cluster,
$v$ does not have edges to $u_1, \ldots,u_{c'}$. 
Therefore,
more than $\{v,w\}$,
each pair of edges $\{v,w_i\}$ and $\{w_i,w\}$ implies
the existence of a certain number of paths between $v$ and $w$ in the original graph.
However, since $v$ is not connected to the neighbors of $u$,
pairs $\{v,u_i\}$ and $\{u_i,u\}$ do not imply 
the existence of a certain number of paths between $v$ and $u$ in the original graph.
Thus, it is likely that $v$ and $w$ that belong to the same cluster
in the transitivity graph, have much more paths between them in the original graph
than $v$ and $u$ that do not belong to the same cluster
in the transitivity graph.


\subsection{Loss function}
\label{Loss function}

In graph neural networks, using both supervised and unsupervised loss functions is common.
In a supervised loss function, nodes' real labels are available and embeddings are learned
such that nodes real labels become similar to their predicted labels,
obtained using the learned embeddings. 
In an unsupervised loss function, embeddings are learned so that nodes that are similar to each other in the graph, find similar embeddings in the vector space.
Formally, the supervised loss function of graph $G$ is defined as follows: 
\begin{equation}
	\label{eg:loss}
	\pounds_{G}^{sup}= -\sum_{v\in V_T(G)}\sum_{c=1}^C l_{v,c}\log(y_{v,c}),
\end{equation}
where $V_T(G)$ is the set of training nodes in graph $G$, $C$ is the number of classes, $l_{v,c}$ is the real label (probability) of belonging node $v$ to class $c$ and $y_{v,c}$ is the predicted label (probability), according to the structure  of $G$.
Similarly, we define the supervised loss function of the transitivity graph $G'$ as follows:
\begin{equation}
	\label{eq:lossc}
	\pounds_{G{'}}^{sup} =	-\sum_{v\in V_T(G{'})}\sum_{c=1}^Cl_{v,c}\log(y'_{v,c}),
\end{equation}
where $V_T(G')$ is the set of training nodes in $G'$ and
$y'_{v,c}$ is the predicted label (probability),
according to the structure  of $G'$. 

In order to devise a unified loss function that effectively accounts for both the relationships within the original graph and those revealed by the transitivity graph, which signifies strong transitivity in the original graph, we introduce the following general loss function:
\begin{equation}
	\label{eq:lossT}
	\pounds^{sup}=	AGG_{\pounds}(\pounds_{G}^{sup}\ ,\pounds_{G'}^{sup}),
\end{equation}
where $AGG_{\pounds}$ is an aggregation operator,
aggregating the two loss functions.
In this paper, we use sum as the aggregation operator of loss functions.
As a result, we obtain the following supervised loss function:
\begin{equation}
	\label{eq:lossT+}
	\pounds^{sup}=	\pounds_{G}^{sup} + \pounds_{G'}^{sup}.
\end{equation}

The unsupervised loss function is formally defined as follows:
\begin{equation}
	\label{eg:unloss}
	\pounds_{G}^{unsup}= -\sum_{v\in V_T(G)}\sum_{u\in N_G(v)} - \log( \sigma(\textbf{z}_v^T \textbf{z}_u) - \sum_{q=1}^Q \mathbb{ E}_{{v_q}\sim P_{q}(v)}\log( \sigma(-\textbf{z}_v^T \textbf{z}_{v_q})),
\end{equation}
where $N_G(v)$ is
defined as the set of nodes that are considered to be similar to $v$ in $G$.
For example, it can be the set of nodes that are connected to $v$ in $G$,
or the set of nodes that are met during some fixed-length random walks starting from $v$. 
Vector $\textbf{z}_v$ is the embedding of $v$, $\sigma$ is the sigmoid function,
$P_q$ is the negative sampling distribution over nodes of $G$ and $Q$ is the number of negative samples.
In a similar way, the unsupervised loss function of transitivity graph $G'$ is defined as follows:
\begin{equation}
	\label{eq:unlossc}
	\pounds_{G'}^{unsup}= -\sum_{v\in V_T(G')}\sum_{u\in N_{G'}(v)} - \log( \sigma(\textbf{z}_v^T \textbf{z}_u) - \sum_{q=1}^Q \mathbb{ E}_{{v_q}\sim P'_{q}(v)}\log( \sigma(-\textbf{z}_v^T \textbf{z}_{v_q})),
\end{equation}
where $N_{G'}(v)$ is
the set of nodes that are similar to $v$ in $G'$ and
$P'_q$ is the negative sampling distribution over nodes of $G'$. 
Similar to the case of supervised loss function,
we define the general unsupervised loss function, as follows:
\begin{equation}
	\label{eq:unsuplossT}
	\pounds^{unsup}=	\pounds_{G}^{unsup} + \pounds_{G'}^{unsup}.
\end{equation} 

Note that on the one hand, $\pounds_{G}^{unsup}$ tries to compute similar embeddings
for nodes that are connected to each other in the original graph.
On the other hand, it aims to compute similar  embeddings
for nodes that are connected to each other in the transitivity graph and as a result,
have strong transitivity relations in the original graph.
In this way, both local connections and strong transitivity relations
are captured by $\pounds_{G}^{unsup}$.
Using reasonable assumptions, similar arguments can be developed for the 
functionality of the supervised loss function.

\subsection{Our proposed model}
\label{Proposed model}

Our proposed model utilizes the aforementioned loss functions to generate embeddings specifically tailored for downstream tasks. Given that we are working with two distinct graphs – the original graph and the transitivity graph – we require a bipartite GNN capable of effectively handling both graphs and generating meaningful embeddings. Motivated by the fusion of Siamese networks \cite{bromley1993signature} and graph matching networks\cite{li2019graph}, we introduce a novel bipartite graph neural network architecture with shared weights. Siamese networks are renowned for their ability to learn similarity between pairs of inputs, while graph matching networks excel at handling complex matching tasks. By incorporating both concepts, our proposed approach aims to leverage their strengths and create a more powerful and versatile network.

Our proposed bipartite graph neural network consists of two parallel branches, each dedicated to processing one input from the pair of graphs $G$ (the original graph) and
$G'$ (the transitivity graph).
The objective of these branches is to extract meaningful embeddings or representations that capture the essential characteristics of the respective inputs.
To accomplish this, the bipartite graph neural network
leverages shared weights, ensuring that the same set of weights is utilized across both branches. This weight sharing mechanism encourages the network to learn and emphasize common patterns/not common patterns and similarities/dissimilarities between the input objects.
Our relations to update nodes' embeddings are as follows:

\begin{equation}
	\label{eq:mesg}
	h_{(G,i)}^{l+1} = \sigma\left(\sum_{j\in N_G(i)}W^l h_{(G,j)}^l + b^l\right)
\end{equation}
\begin{equation}
	\label{eq:mesc}
	h_{(G',i)}^{l+1} = \sigma\left(\sum_{j\in N_{G'}(i)}W^l h_{(G',j)}^l + b^l\right),
\end{equation}
where $h_{(G,i)}^{l}$ and $h_{(G',i)}^{l}$ are respectively layer-$l$ embeddings of node $i$ in graphs $G$ and $G'$, and $W^l$
and $b^l$ are trainable parameters.
As the input graphs $G$ and $G'$ are processed through their respective branches, each branch produces an embedding that captures the salient features of the corresponding object.
By utilizing shared weights and processing the inputs in parallel, our bipartite graph neural network can effectively capture and encode the inherent similarities/real connections and strong transitivity between the objects in the pair.
Throughout the training process, the weights of the bipartite graph neural network are updated using
the loss function in Equation \ref{eq:lossT},
which is simultaneously
associated with both the original graph and
the transitivity graph.

\begin{algorithm}
	\caption{Transitivity-aware GNN}\label{alg1}
	\hspace*{\algorithmicindent} {Input: Graph $G(V,E,X)$, the number of layers $L$}. \\
	\hspace*{\algorithmicindent} {Output: Matrix $H$ whose $i$-th row includes the embedding of node $i$.}
	\begin{algorithmic}[1]
		\State$H\gets \emptyset$.
		\State $\textit{$G'$} \gets \text{transitivity graph of  }\textit{G}$.
		\State $G'' \gets \textit{cluster $G'$}$.
		\For{ $l:1 \ to \ L$}
		\For{  $i \ in \ V$}
		\State $ h_{(G, i)}^{l}\gets \text{update the embedding of node $i$ according to \autoref{eq:mesg}}$.
		\State $ h_{(G'', i)}^{l}\gets\text{update the embedding of node $i$ according to \autoref{eq:mesc}}$.
		\State $H^{i} = h_{(G, i)}^{l}$.
		\EndFor
		\EndFor
		\State $\text{return $H$}$.
		\end{algorithmic}
\end{algorithm}

Algorithm \ref{alg1} presents the high level pseudocode of our node embedding generation method, called Transitivity GNN, or TransGNN in short.
It first creates the transitivity graph from the original graph.
Then it feeds both the original graph and its transitivity graph into
a graph neural network, following the principles elucidated in Equations \ref{eq:mesg} and \ref{eq:mesc}.
The SSC condition of strong transitivity relations is satisfied in Line 2,
where in $G'$ we selectively preserve those edges that the structural similarity of their endpoints in the original graph exceeds a user-defined threshold.
To satisfy the SNPC condition, we apply a graph clustering algorithm,
as depicted in Line 3 of Algorithm \ref{alg1}.
In Lines 6 and 7 of Algorithm \ref{alg1},
we learn embeddings for both the original graph and the clustered transitivity graph,
where the trainable weight parameters are shared.
However, we return the embeddings learned for the nodes of $G$, as the final embeddings.
In other words, the impact of learning over the transitivity graph is
to improve the weights of the model that learns embeddings for the original graph.
The learned embeddings can be used for a downstream task, such  as node classification.

\section{Experiments}\label{Experiments}

Given that our proposed model is adaptable to different GNN models, our evaluation involves comparisons between a number of well-known GNN models,
with and without improving them by our method.
To do so, we analyze the performance of GCN \cite{kipf2016semi}, GAT \cite{velickovic2017graph}, GATv2 \cite{brody2021attentive} and SGC  \cite{wu2019simplifying}, with and without the application of our method.

\subsection{Datasets}

We use the following famous datasets in our comparisons:
 \paragraph{Citation networks}\footnote{\href{https://github.com/kimiyoung/planetoid/tree/master/data}{https://github.com/kimiyoung/planetoid/tree/master/data}} \ We examine three citation networks--Cora, Citeseer, and Pubmed--following the work of Sen et al. \cite{sen2008collective}. These datasets comprise sparsely represented bag-of-words feature vectors for each document, accompanied by a roster of citation connections among the documents.

\paragraph{United States Air-traffic network}\footnote{\href{https://github.com/leoribeiro/struc2vec/tree/master/graph}{https://github.com/leoribeiro/struc2vec/tree/master/graph}} \ This data is compiled from the bureau of transportation statistics, spanning January to October 2016. Airport activity is quantified by the cumulative count of individuals who transited (including arrivals and departures) through the airport during the corresponding timeframe \cite{ribeiro2017struc2vec}. Node attributes are represented using one-hot encoding of the respective nodes \cite{jin2020gralsp}.

\paragraph{Twitch social}\footnote{\href{https://graphmining.ai/datasets/ptg/twitch/}{https://graphmining.ai/datasets/ptg/twitch/}} \ This dataset includes networks of user interactions in which nodes represent Twitch users connected by mutual friendships. Node attributes encompass preferred games, geographic location, and streaming patterns \cite{rozemberczki2021multi}. Within this dataset, there are six distinct graph collections: DE, EN, ES, FR, PT, and RU. However, we exclusively focus on the PT graph for our evaluation .

\paragraph{Network of Actor Co-occurrences}\footnote{\href{https://github.com/bingzhewei/geom-gcn/tree/master/new_data}{https://github.com/bingzhewei/geom-gcn/tree/master/new\_data}} \  This dataset presents an actor-centric subset extracted from the film-director-actor-writer network \cite{pei2020geom}. Each node represents an actor, with edges connecting nodes that co-occur on Wikipedia pages. Node attributes capture specific keywords from Wikipedia entries. We categorize nodes into five classes based on the terminology found in the corresponding actor's Wikipedia content.

Table \ref{table:1} summarizes the statistics of our used datasets.

\begin{table}[h!]
	\centering	
	\caption{Dataset statistics.\label{table:1}}
	\begin{tabular}{l*{4}{c}|c}
		\hline
		Dataset              & nodes & edge & feature & class & Train/Dev/Test nodes \\
		\hline
		Cora & 2708 & 10556 & 1433 & 7 & 140/500/1000  \\
		Citeseer & 3327 & 9104 & 3703 & 6& 120/500/1000   \\
		Pubmed & 19717 & 88648 & 500 & 3  & 60/500/1000   \\
		Airport(usa) & 1190 & 13599 & 1190 & 4&65/119/595   \\
		Actor & 7600 & 30019 & 932 & 5 & 418/760/3800  \\
		Twitch  PT & 1912 & 64510 & 128 & 2 & 105/191/956 \\
		\hline	
	\end{tabular}		
\end{table}

\begin{table}[h!]
	\centering
	\caption{The number of edges in the transitivity graphs of our used  datasets.\label{table:2}}	
	\begin{tabular}{l*{3}{c}r}
		\hline
		Dataset               & $\#$ edge &   Threshold \\
		\hline
		Cora  & 2610 & 0.4  \\
		Citeseer &  3554 &  0.5  \\
		Pubmed & 80498 &  0.4  \\
		Airport(usa)  & 1496 & 0.5  \\
		Actor & 39844 &0.88  \\
		Twitch  PT  & 2158 & 0.09   \\
		\hline
	\end{tabular}		
\end{table}

\subsection{Implementation details}
\label{sec:imp}

To implement our TransGNN model, which entails incorporating a transitivity graph into existing GNN models, we follow the subsequent steps:

\paragraph{Step 1} To acquire the transitivity graph, specifically the SSC condition of strong transitivity relations, we calculate node similarities using the SimRank algorithm \cite{jeh2002simrank}. Then, we construct a transitivity graph where nodes
that are unconnected to each other in the original graph but expose a similarity value higher than a predefined threshold, are connected (Line 2 of Algorithm \ref{alg1}).  
Table \ref{table:2} displays for each dataset the corresponding threshold value, which is determined through a hyperparameter tuning process using a greedy search.
It also depicts the number of edges that the transitivity graph finds in this way.
Note that in terms of statistics,
the sole difference between the transitivity graph and the original graph lies in the number of their edges; the other attributes such as the number of nodes, features, and classes remain the same.
A noteworthy observation in Table \ref{table:2} is that the number of
edges in the transitivity graph closely aligns with the number of edges in the original graph.

\paragraph{Step 2} As previously discussed, we employ a graph clustering algorithm to fulfill the SNPC condition. we employ the Metis algorithm \cite{karypis1998fast}, drawing inspiration from the approach used in \cite{chiang2019cluster}. The Metis algorithm aims to minimize the number of edges between clusters.


Every model comprises two GNN layers and as stated in Section \ref{Loss function}, we opt for the cross-entropy loss function \cite{stone2015information}.
Moreover, we use addition as the aggregation operator of the loss functions \ref{eq:lossT+} and \ref{eq:unsuplossT}.
We set the learning rate and hidden dimension to $0.01$ and $32$, respectively, and employ the Adam optimizer \cite{kingma2014adam}.
For the GAT and GATv2 models, we set the number of attention heads to $8$.
For the SGN model, we set the parameter $K$ to $1$ \cite{wu2019simplifying}. Our parameters configuration aligns with the setting outlined in the work of Yang et al \cite{yang2016revisiting}.

 \subsection{Results}
 
 We run each model for $100$ times and report the average results
 (along with the standard deviations) in Tables \ref{tabel:3} and \ref{table: 4}.
Table \ref{tabel:3} displays the accuracy of the models,
while Table \ref{table: 4} presents the weighted F1 scores.
To compute the weighted F1 score, we calculate precision and recall for each class and compute their average by weighting them with support, which represents the number of true instances of the class.
In Tables \ref{tabel:3} and \ref{table: 4}, each GNN model that
incorporates our proposed method is denoted with "TransGNN", while its original form without using our method is denoted with "GNN".
For instance,
while "SGC" indicates the original SGC model without our proposed enhancement,
"TransSGC" denotes the enhanced SGC model using our proposed transitivity graph technique.
 
\begin{table}[h!]
	\centering
	\caption{Accuracy comparison of GNN models improved by our transitivity graph technique, against their original forms. In each case, the better model is highlighted in bold.\label{tabel:3}}
	\resizebox{\textwidth}{!}{%
	\begin{tabular}{l*{6}{c}r}
		Model              & Cora & Citeseer & Pubmed &  Airport(USA)& Actor & Twitch  PT  \\
		\hline		
		GCN & 81.1 $\pm$ 0.2 & 69.9 $\pm$ 0.2 & 79.3 $\pm$ 0.1 & 38.1 $\pm$ 0.4&24.7 $\pm$ 0.15&61.4 $\pm$ 0.3   \\
		TransGCN &\B 85.1 $\pm$ 0.3 &\B 74.1 $\pm$ 0.12 &\B 80.7 $\pm$ 0.2 &\B 41.8 $\pm$ 0.6&\B 28.2 $\pm$ 0.4&\B 63.7  $\pm$ 0.17  \\
		\hline
		GAT & 82.0 $\pm$ 0.6 & 67.5 $\pm$ 0.7 & 78.1  $\pm$ 0.9 & 46.0  $\pm$ 0.5& 25.2  $\pm$ 0.6& 63.3 $\pm$ 0.7     \\
		TransGAT &\B 84.3 $\pm$ 0.15 &\B 72.7 $\pm$ 0.4 &\B 81.4 $\pm$0.5 &\B 49.2 $\pm$ 0.6&\B 29.9 $\pm$ 0.4 &\B 64.2 $\pm$ 0.42   \\
		\hline
		GATv2 & 79.8 $\pm$ 0.4 & 68.2 $\pm$ 0.22 & 77.9 $\pm$ 0.35 & 47.5 $\pm$ 0.12 & 25.2 $\pm$0.1& 63.3 $\pm$ 0.37  \\
		TransGATv2 &\B  83.4 $\pm$ 0.8 &\B 72.9$\pm$ 0.54 &\B 78.9 $\pm$ 0.2&\B 52.6 $\pm$ 0.22&\B 29.7 $\pm$0.25 &\B 67.1 $\pm$ 0.33 \\
		\hline
		SGC & 81.2 $\pm$ 0.1 & 70.1 $\pm$ 0.12 & 78.7 $\pm$ 0.3 & 38.8 $\pm$ 0.17 & 24.7 $\pm$ 0.1 & 61.5  $\pm$ 0.15   \\
		TransSGC &\B 84.3 $\pm$ 0.3 &\B 75.0 $\pm$ 0.15 &\B 81.3 $\pm$ 0.1 &\B 44.4 $\pm$ 0.15 &\B 27.6 $\pm$0.1 &\B 63.5 $\pm$ 0.2   \\
		\hline
	\end{tabular}}
\end{table}

\begin{table}[h!]
	\centering
	\caption{Comparing weighted F1 scores of the enhanced GNN models, against the original forms. In each case, the better model is highlighted in bold.\label{table: 4}}
	\resizebox{\textwidth}{!}{%
	\begin{tabular}{l*{6}{c}r}
		model              & Cora & Citeseer & Pubmed &  Airport(usa)& Actor & Twitch  PT  \\
		\hline 		
		GCN & 0.811 $\pm$ 0.01 & 0.703 $\pm$ 0.03 & 0.789 $\pm$ 0.01 & 0.326 $\pm$ 0.05&0.239 $\pm$ 0.01& 0.596 $\pm$ 0.02   \\
		TransGCN &\B 0.851 $\pm$ 0.02 &\B 0.742 $\pm$ 0.025 &\B 0.806 $\pm$ 0.01 &\B 0.386 $\pm$ 0.03&\B 0.281 $\pm$ 0.02&\B 0.613 $\pm$ 0.02   \\
		\hline
		GAT & 0.823 $\pm$ 0.03 & 0.690 $\pm$ 0.05 & 0.780 $\pm$ 0.035 & 0.385 $\pm$ 0.032& 0.156 $\pm$ 0.01& 0.533 $\pm$ 0.05     \\
		TransGAT &\B 0.845 $\pm$ 0.06 &\B 0.723 $\pm$ 0.03 &\B 0.801 $\pm$ 0.02 &\B 0.501 $\pm$ 0.025&\B 0.214 $\pm$ 0.03&\B 0.674 $\pm$ 0.03   \\
		\hline
		GATv2 & 0.811 $\pm$ 0.01 & 0.689 $\pm$ 0.03 & 0.778 $\pm$ 0.03 & 0.431 $\pm$ 0.05 & 0.241 $\pm$ 0.04& 0.491 $\pm$ 0.05   \\
		TransGATv2 &\B 0.831 $\pm$ 0.03 &\B 0.728 $\pm$ 0.07 &\B 0.786 $\pm$ 0.02 &\B 0.498 $\pm$ 0.03 &\B 0.263 $\pm$ 0.04 &\B 0.633 $\pm$ 0.02 \\
		\hline
		SGN & 0.812 $\pm$ 0.02 & 0.693 $\pm$ 0.01 & 0.786 $\pm$ 0.01 & 0.324 $\pm$ 0.01 & 0.232 $\pm$ 0.01 & 0.590 $\pm$ 0.02   \\
		TransSGN &\B 0.850 $\pm$ 0.02 &\B 0.750 $\pm$ 0.01 &\B 0.812 $\pm$ 0.01 &\B 0.466 $\pm$ 0.05 &\B 0.259 $\pm$ 0.02 &\B 0.643 $\pm$ 0.035   \\
		\hline 		
	\end{tabular}}
\end{table}

In Tables \ref{tabel:3} and \ref{table: 4}, for each GNN model
and over each dataset, between the original GNN and its enhanced form,
the one that yields a better result is highlighted in bold. 
It is evident from the tables that our proposed method has led to significant performance improvements in the basic models.
These enhancements are notable,
ranging from $1\%$ to $14\%$ across the various datasets we assessed.
This improvement highlights the efficacy of our proposed approach in bolstering the capabilities of various GNN models, across diverse data scenarios.
 
 \subsection{Robustness analysis}
 \label{sec:robustness}

\begin{figure}[h!]
 	\centering
 	\begin{subfigure}[b]{0.4\linewidth}
 		\includegraphics[angle=0,width=\linewidth]{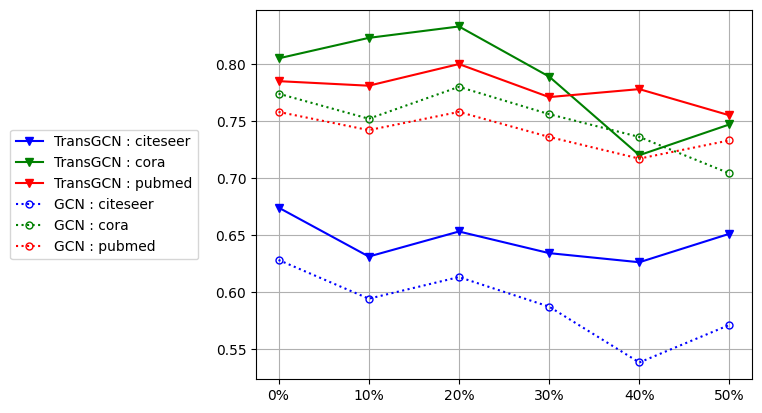}
 		\caption{removing edges - GCN}
 		
 	\end{subfigure}
 	\begin{subfigure}[b]{0.4\linewidth}
 		\includegraphics[angle=0,width=\linewidth]{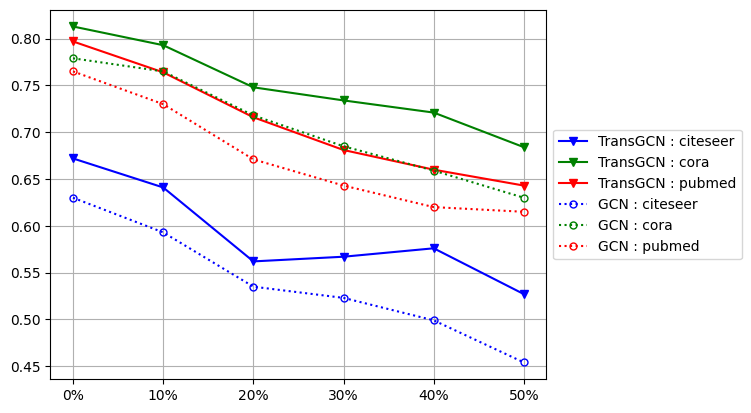}
 		\caption{adding edges - GCN}
 	\end{subfigure}
 	\begin{subfigure}[b]{0.4\linewidth}
 		\includegraphics[angle=0,width=\linewidth]{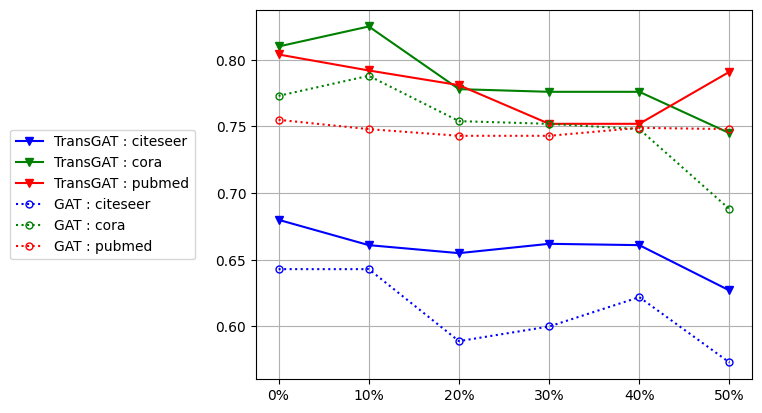}
 		\caption{removing edges - GAT}
 		
 	\end{subfigure}
 	\begin{subfigure}[b]{0.4\linewidth}
 		\includegraphics[angle=0,width=\linewidth]{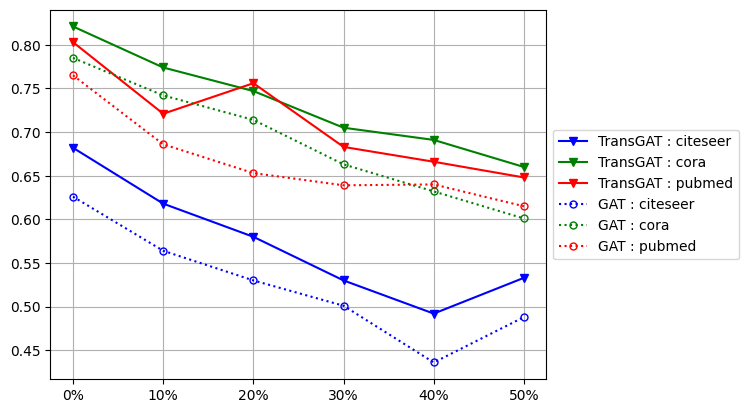}
 		\caption{adding edges - GAT}
 	\end{subfigure}
 	\begin{subfigure}[b]{0.4\linewidth}
 		\includegraphics[angle=0,width=\linewidth]{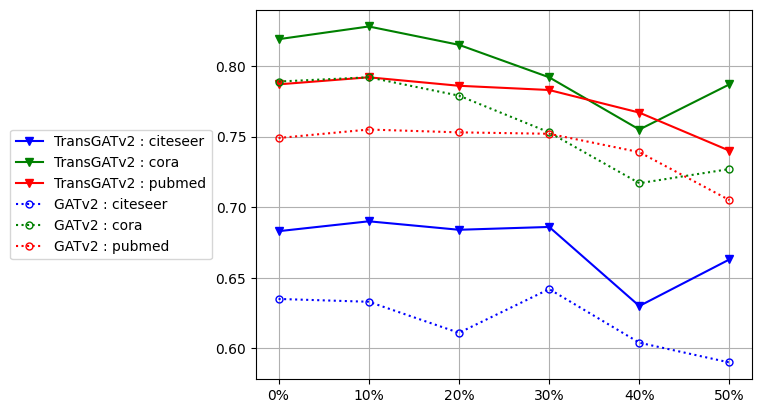}
 		\caption{removing edges - GATv2}
 		
 	\end{subfigure}
 	\begin{subfigure}[b]{0.4\linewidth}
 		\includegraphics[angle=0,width=\linewidth]{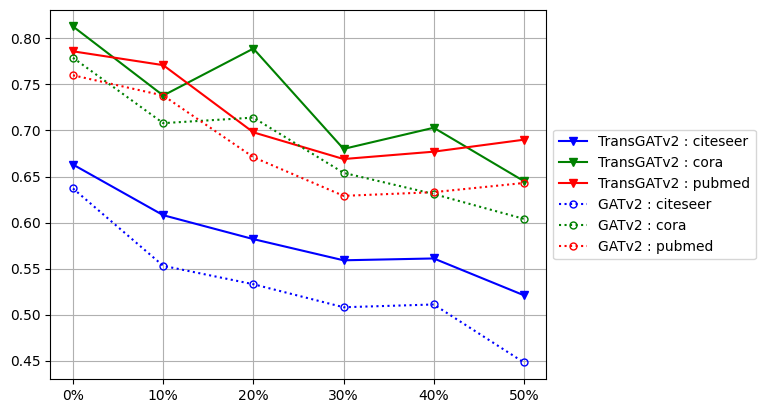}
 		\caption{adding edges - GATv2}
 	\end{subfigure}
 	\begin{subfigure}[b]{0.4\linewidth}
 		\includegraphics[angle=0,width=\linewidth]{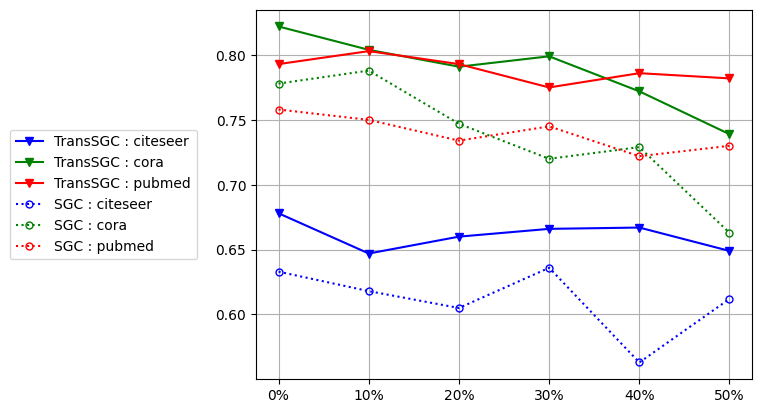}
 		\caption{removing edges - SGC}
 		
 	\end{subfigure}
 	\begin{subfigure}[b]{0.4\linewidth}
 		\includegraphics[angle=0,width=\linewidth]{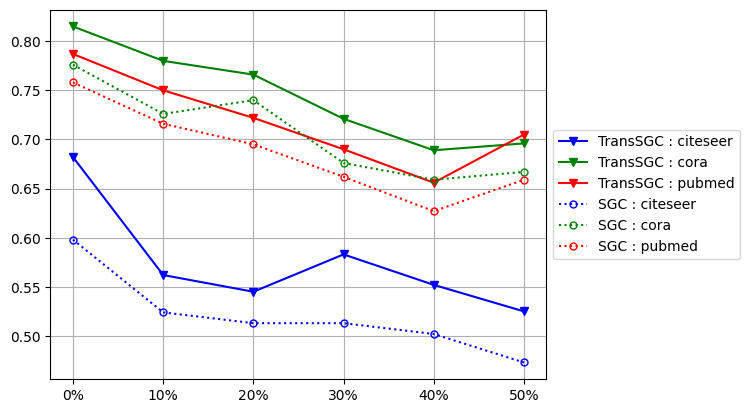}
 		\caption{adding edges - SGC}
 	\end{subfigure}
 	\caption{GNNs enhanced with our proposed method are more resilient to noise (edge addition and edge removal), than the basic GNNs.
 		 The vertical axes show accuracy and the horizontal axes present the percentage of edges added or removed.\label{fig:fig2}}
 \end{figure}

An interesting question about the performance of a GNN model is its
robustness against noise.
In this section, 
we examine how robust is our model against noises in the forms of 
edges added and edges removed from the input graph.
More precisely, by adding some edges
we induce noise to the input graph, train the model on this noisy graph
and test the model on the original graph.
We perform the same procedure with removing some edges from the input graph. 
We conduct this robustness analysis over the three citation networks.
We examine different number of edges added/removed,
ranging from $0\%$ to $50\%$ of the number of original edges in the graph. 
As depicted in Figure \ref{fig:fig2},
the models enhanced with our proposed technique show a better resilience than the basic models, in terms of accuracy, throughout the process of adding or removing edges.
Additionally, the enhanced models showcase less performance fluctuation compared to the basic models, across most of the datasets.

In general in our experiments, adding edges shows more fluctuations than
 removing edges.
A reason for this phenomenon 
can be the way we construct the transitivity graph:
the input graph and its transitivity are mutually exclusive, with the edges present in the input graph being absent in the transitivity graph, and vice versa.
Adding edges to the input graph limits the number of potential choices for the transitivity graph. Therefore, it increases the likelihood of having shared edges between the transitivity of the original graph and the transitivity of the noisy graph. However, as discussed in Section \ref{transitive graph}, it's important to note that the original graph and the transitivity graph should not share any edges in common.
Inversely, removing edges from the original graph expands the spectrum of choices available to the transitivity graph, which mitigates the potential of having shared edges between the transitivity of the original graph and the transitivity of the noisy graph.

\subsection{Ablation study}
\label{sec:ablation}
 
In this section, we examine the impact of using different loss functions and assess how they affect the performance of our proposed model.
In our analysis, we consider the following loss functions: 
\begin{itemize}
\item 
embedding-associated loss function:
the loss function used by an original GNN model to compute embeddings based on the original graph. We refer to it as "loss".
\item 
transitivity loss function:
the loss function used over the transitivity graph. We refer to it as "trans\_loss".
\item 
similarity-based loss function: 
this loss function is defined as $ 1- cosine\ similarity$ between embeddings from the original graph and the transitivity graph. We refer to as "sim\_loss".
The intuition behind this loss function is to heighten and emphasize the similarity/difference between the embeddings of the original graph and the embeddings of the transitivity graph.
Note that $ 1- cosine\ similarity$ measures the amount of dissimilarity/distance
between the embeddings.
By incorporating 'sim\_loss', our proposed model becomes more adept at capturing the intuitive similarity between the original graph and the transitivity graph and as a result, nodes' similarities modeled by the transitivity graph. By employing this loss function, we can effectively account for various levels of similarity within the original graph, as previously discussed.
\end{itemize}

\begin{figure}[h!]
	\centering
	\begin{subfigure}[b]{0.3\linewidth}
		\includegraphics[angle=0,width=\linewidth]{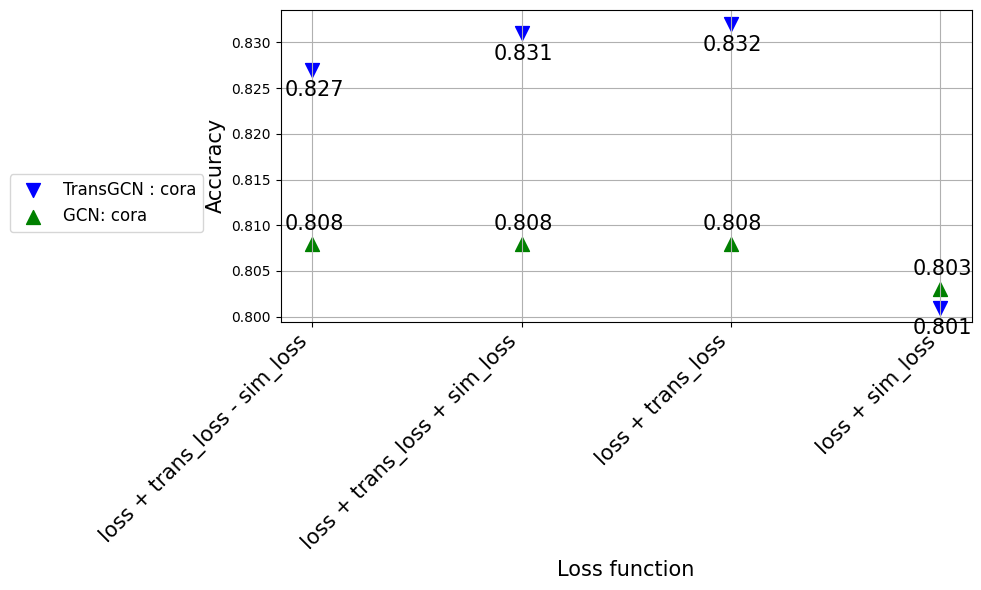}
		\caption{}
		\
	\end{subfigure}
	\begin{subfigure}[b]{0.3\linewidth}
		\includegraphics[angle=0,width=\linewidth]{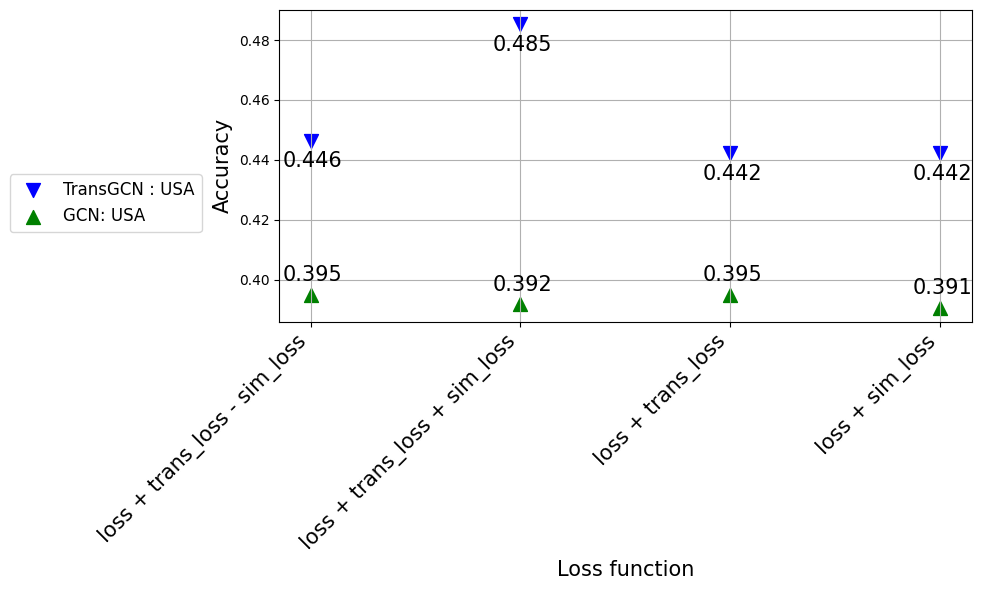}
		\caption{}
		\
	\end{subfigure}
	\begin{subfigure}[b]{0.3\linewidth}
		\includegraphics[angle=0,width=\linewidth]{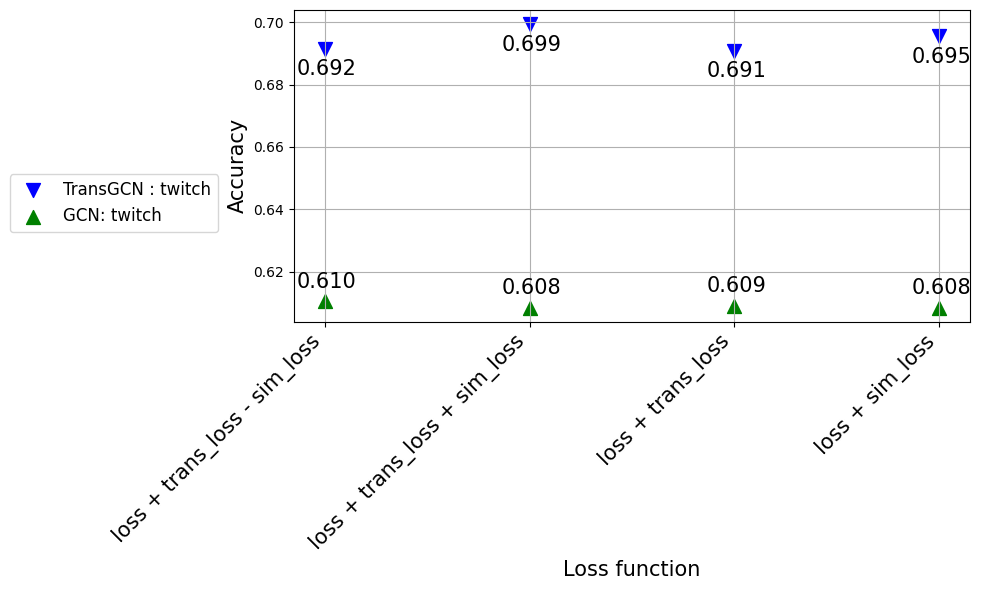}
		\caption{}
		\
	\end{subfigure}
	
	\begin{subfigure}[b]{0.3\linewidth}
		\includegraphics[angle=0,width=\linewidth]{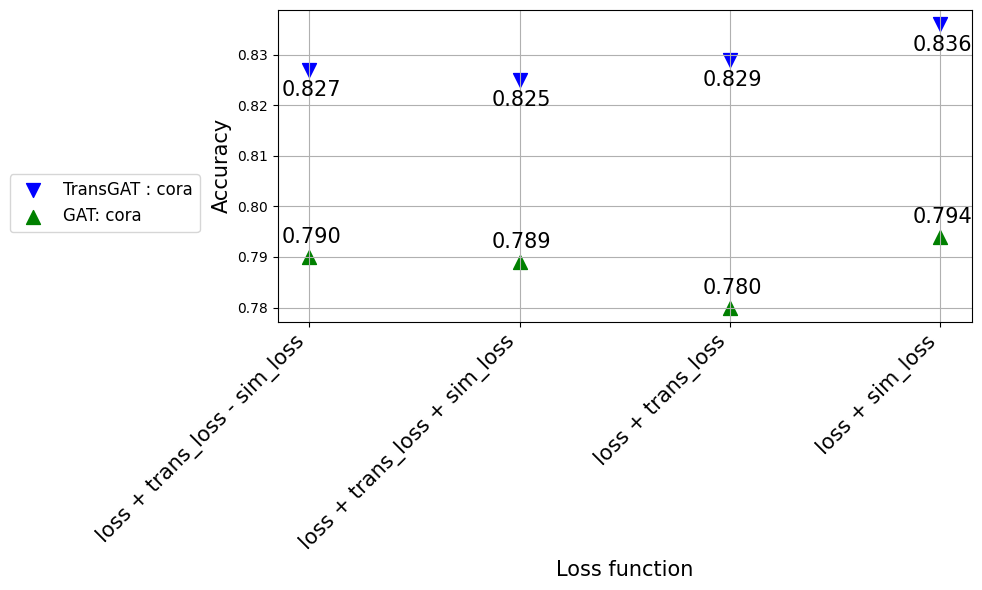}
		\caption{}
		\
	\end{subfigure}
	\begin{subfigure}[b]{0.3\linewidth}
		\includegraphics[angle=0,width=\linewidth]{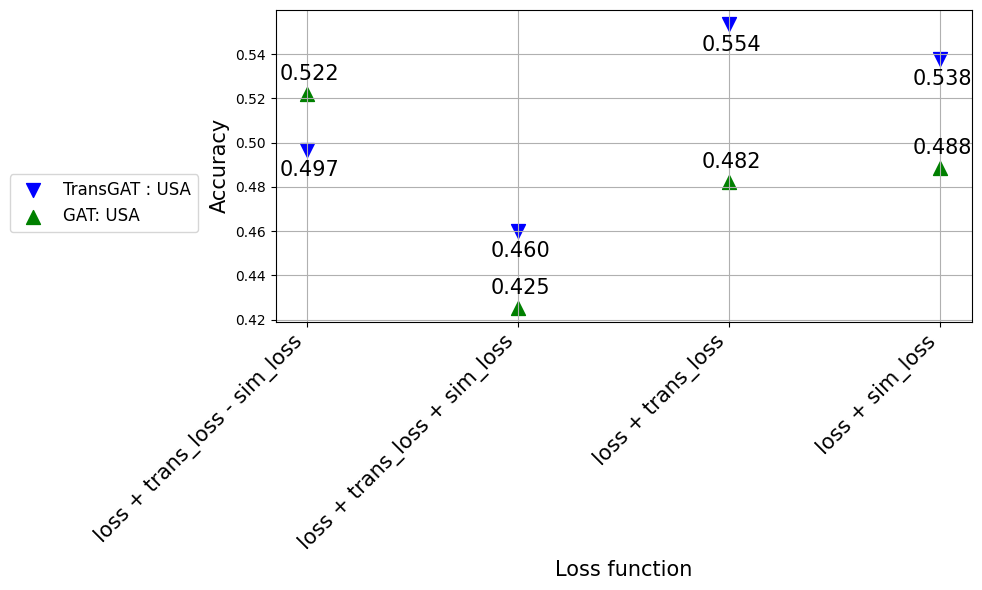}
		\caption{}
		\
	\end{subfigure}
	\begin{subfigure}[b]{0.3\linewidth}
		\includegraphics[angle=0,width=\linewidth]{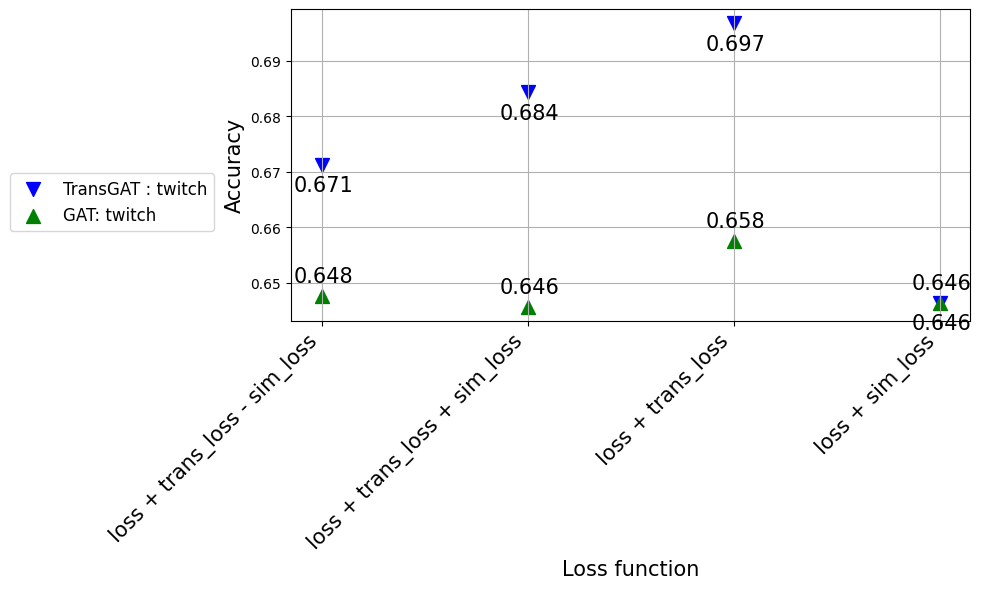}
		\caption{}
		\
	\end{subfigure}
	
	\begin{subfigure}[b]{0.3\linewidth}
		\includegraphics[angle=0,width=\linewidth]{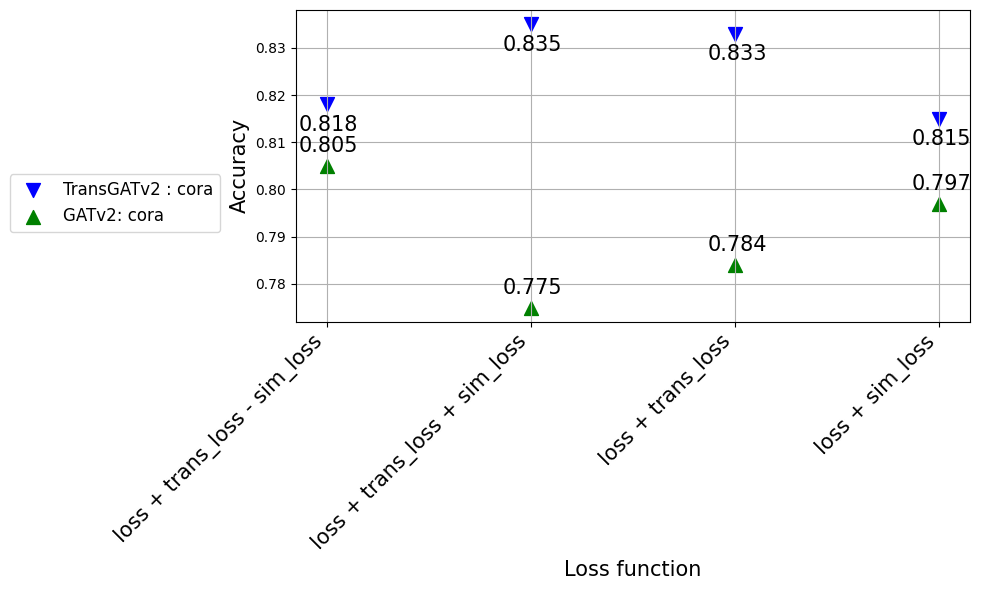}
		\caption{}
		\
	\end{subfigure}
	\begin{subfigure}[b]{0.3\linewidth}
		\includegraphics[angle=0,width=\linewidth]{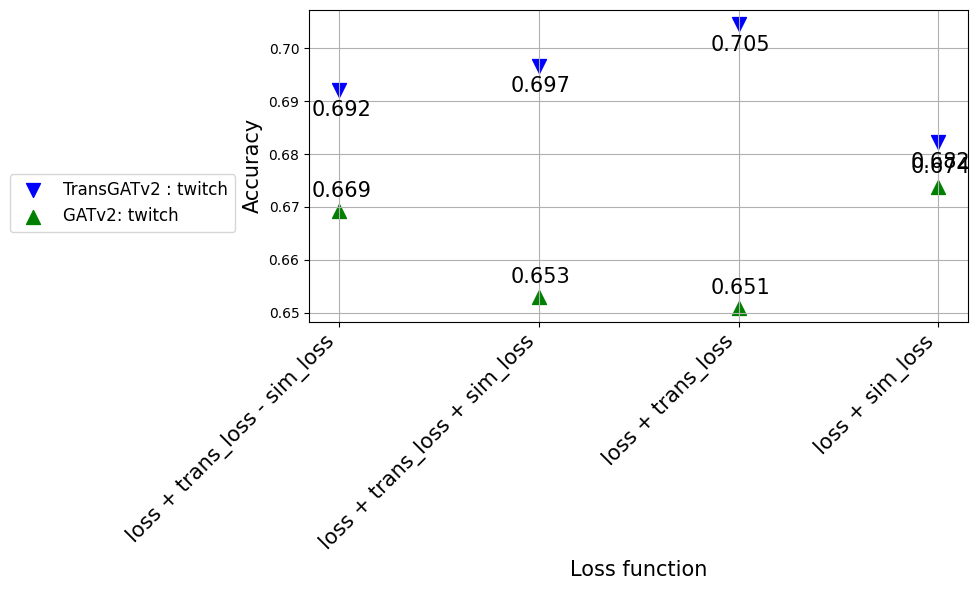}
		\caption{}
		\
	\end{subfigure}
	\begin{subfigure}[b]{0.3\linewidth}
		\includegraphics[angle=0,width=\linewidth]{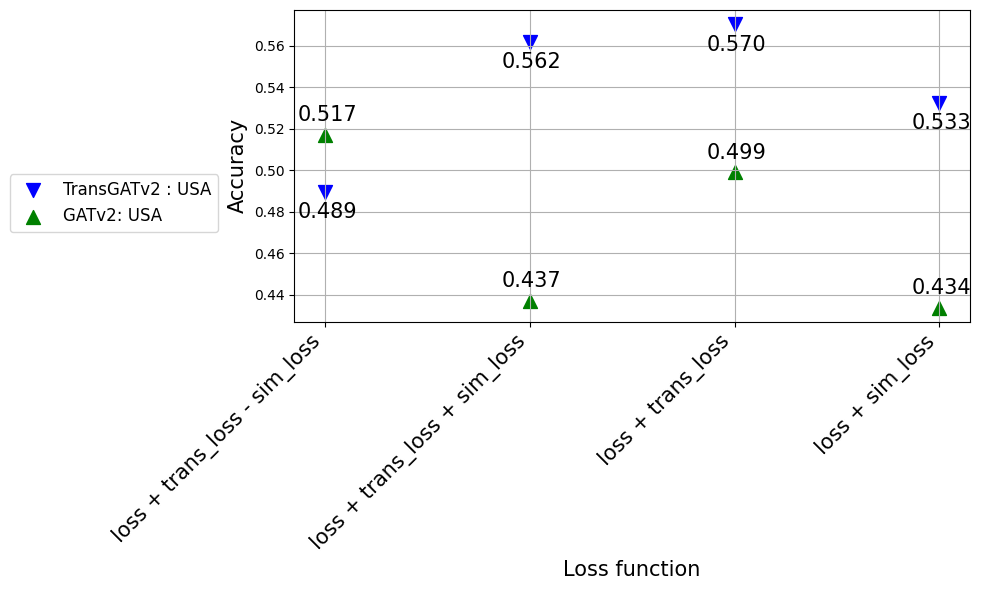}
		\caption{}
		\
	\end{subfigure}
	
	\begin{subfigure}[b]{0.3\linewidth}
		\includegraphics[angle=0,width=\linewidth]{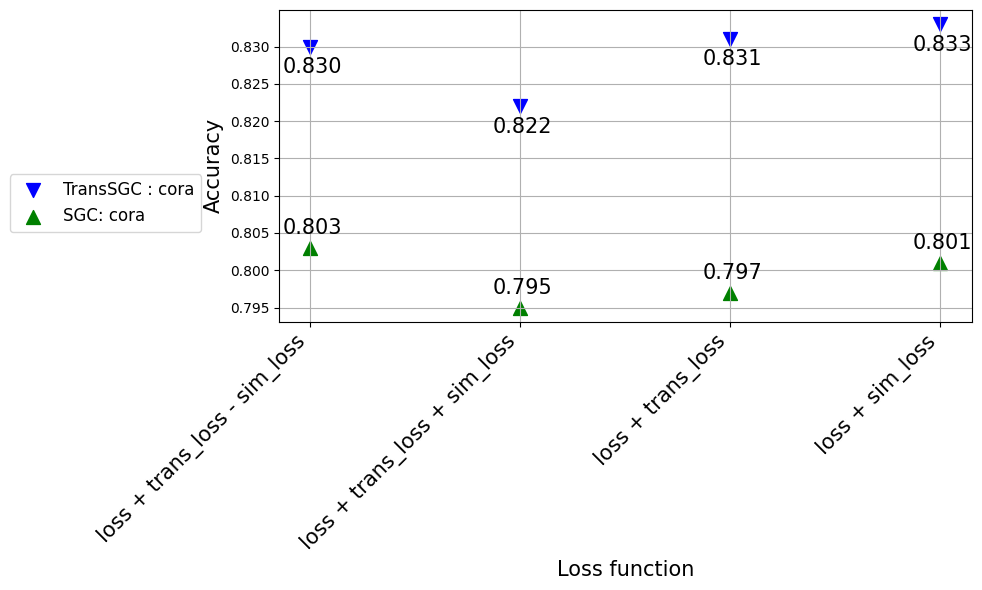}
		\caption{}
		\
	\end{subfigure}
	\begin{subfigure}[b]{0.3\linewidth}
		\includegraphics[angle=0,width=\linewidth]{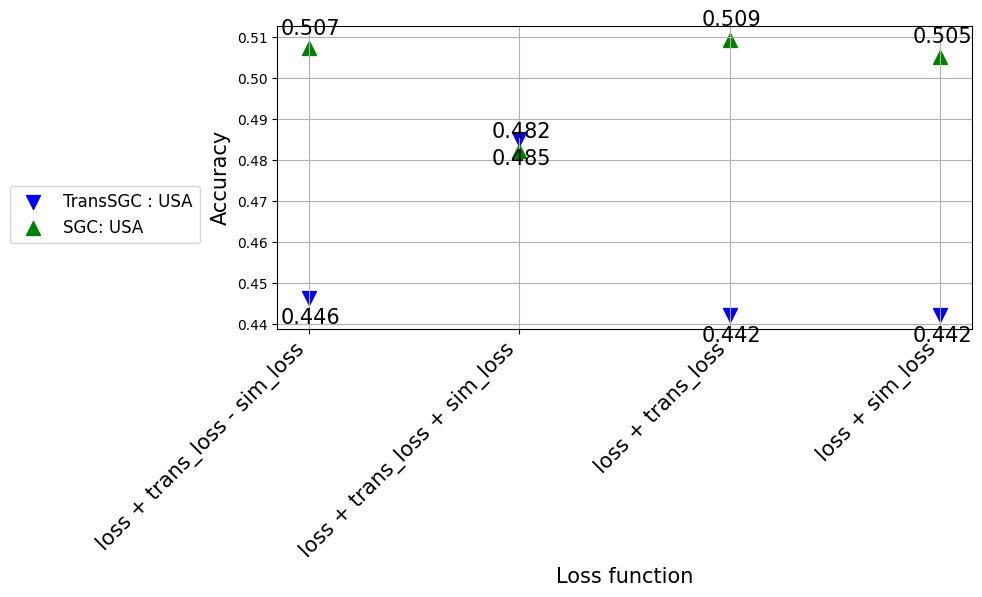}
		\caption{}
		\
	\end{subfigure}
	\begin{subfigure}[b]{0.3\linewidth}
		\includegraphics[angle=0,width=\linewidth]{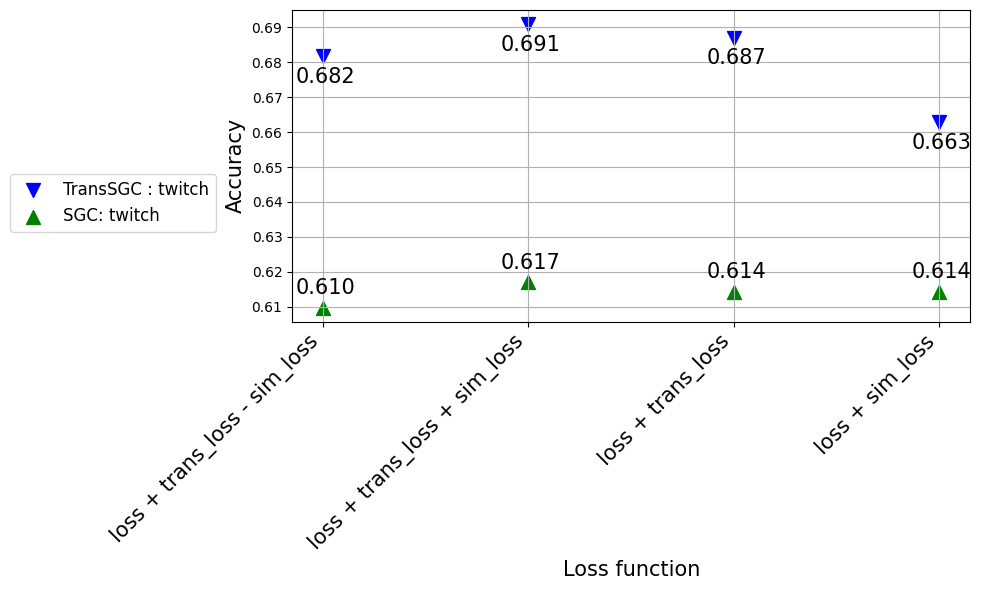}
		\caption{}
		\
	\end{subfigure} 	 	
	\caption{Accuracy comparison of our proposed model against the original models across different aggregations of loss functions on Cora, Airport (USA), and Twitch PT.\label{fig:fig3}
In this context, "loss" refers to the loss function associated with the embedding of the original graph, "trans\_loss" refers to the loss related to the embedding of the transitivity graph, and "sim\_loss" deals with the cosine similarity (calculated as $1 - cosine \ similarity$) between the embeddings of the original and transitivity graph.
		}
\end{figure}

Figure \ref{fig:fig3} illustrates the accuracy comparison between our proposed model and the corresponding basic model, across three datasets: Cora, Airport(USA) and Twitch PT. We explore various combinations for the three discussed loss functions. Specifically, we investigate  the effects of the following combinations:
 
\begin{enumerate}
\item 
"loss + trans\_loss - sim\_loss": we observe that this combination consistently outperforms the others, indicating that the joint inclusion of "trans\_loss" and "sim\_loss" in addition to the base function "loss" contribute positively to similarity-level measurement and to the overall accuracy. This implies that "sim\_loss", while not effectively capturing the level of similarities on its own, can enhance the impact of "trans\_loss" when combined with the base loss function.
\item 
"loss + trans\_loss + sim\_loss": Similar to the previous combination, when we add "sim\_loss" to "trans\_loss" and the base loss, it results in better outcomes. This highlights the positive impact of "sim\_loss" in measuring similarity levels when combined with other relevant loss components.
\item 
"loss + trans\_loss": This combination notably enhances the performance of "loss". Nevertheless, the absence of "sim\_loss" in this combination appears to restrict its ability to capture the complete spectrum of similarity levels.
\item 
"loss + sim\_loss": Interestingly, in many scenarios, this combination doesn't yield significant gains over "loss". This suggests that "sim\_loss" on its own might not be a robust indicator of the level of similarities, especially when compared to the benefits derived from the joint usage of "trans\_loss" and "sim\_loss".
\end{enumerate}

In Figure \ref{fig:fig3}, it is important to note that the base models exclusively use "loss", without integrating it with any other loss functions.
This means that in each setting,
we run  both our proposed model and its corresponding base model for several times, where the base model only employs the "loss" function. 
The results reported in Figure \ref{fig:fig3} are the averages of these multiple runs.
In summary, our experiments demonstrate that the integration of "trans\_loss" and "sim\_loss" with the base "loss"  assesses strong transitivity relations in the original graph, more effectively. Consequently, this integration leads to an enhancement in overall accuracy.
This combination not only strengthens our model but also provides a more comprehensive representation of both real connections and strong transitivity relations. Furthermore, it takes into consideration the nuances of similarity levels within the graph, ultimately leading to improved accuracy and performance.

\section{Conclusion}
\label{Conclusion}
 
In this paper, we demonstrated and highlighted the significance of
strong transitivity and its connection with varying degrees of similarity among nodes.
This understanding provides valuable insights for the development of effective embedding learning mechanisms. Subsequently, we introduced the concept of using a transitivity graph derived from the input graph to model and represent strong transitivity relations.
We accordingly developed the Transitivity Graph Neural Network (or TransGNN in short) model, that can be aligned with existing GNNs and improve their similarity-awareness. In this way, both local connections between nodes and strong transitivity relations are captured by a graph neural network, resulting in learning higher-quality embeddings. We examined our model over several real-world datasets and showed that it improves the performance of several well-known GNNs, for tasks such as node classification.

 \bibliographystyle{IEEEtran}
 \bibliography{./cite.bib}
 
\end{document}